\documentclass[twocolumn,superscriptaddress]{revtex4-1}
\pdfoutput=1
\usepackage[bookmarks=false]{hyperref}
\usepackage{dcolumn}
\usepackage{bm}
\usepackage{ifthen}
\usepackage[usenames]{color}
\usepackage{amssymb}
\usepackage{amsfonts}
\usepackage{amsmath}
\usepackage{graphicx}
\usepackage{epstopdf}
\usepackage{xcolor}
\hypersetup{
    colorlinks,
    linkcolor={red!50!black},
    citecolor={blue!50!black},
    urlcolor={blue!80!black}
}
\usepackage{cancel}
\usepackage{soul}
\usepackage{multirow}

\newcommand{\etal}{\textit{et al.}}

\makeatletter
\renewcommand{\p@subsection}{}
\renewcommand{\p@subsubsection}{}
\makeatother

\begin{document}

\title{Compositional optimization of hard-magnetic phases with machine-learning models}

\author{Johannes J. M\"oller}
\email[Corresponding author: ]{johannes.moeller@iwm.fraunhofer.de} 

\author{Wolfgang K\"orner}
\affiliation{Fraunhofer Institute for Mechanics of Materials IWM, W\"ohlerstr. 11, 79108 Freiburg, Germany}\author{Georg Krugel}
\author{Daniel F. Urban}
\affiliation{Fraunhofer Institute for Mechanics of Materials IWM, W\"ohlerstr. 11, 79108 Freiburg, Germany}
\author{Christian Els\"asser}
\affiliation{Fraunhofer Institute for Mechanics of Materials IWM, W\"ohlerstr. 11, 79108 Freiburg, Germany}
\affiliation{University of Freiburg, Freiburg Materials Research Center, Stefan-Meier-Str. 21, 79104 Freiburg, Germany}

\begin{abstract}
Machine Learning (ML) plays an increasingly important role in the discovery and design of new materials.
In this paper, we demonstrate the potential of ML for materials research using hard-magnetic  phases as an illustrative case.
We build kernel-based ML models to predict optimal chemical compositions for new permanent magnets, which are key components in many green-energy technologies. 
The magnetic-property data used for training and testing the ML models are obtained from a combinatorial high-throughput screening based on density-functional theory calculations. 
Our straightforward choice of describing the different configurations enables the subsequent use of the ML models for compositional optimization and thereby the prediction of promising substitutes of state-of-the-art magnetic materials like Nd$_2$Fe$_{14}$B with similar intrinsic hard-magnetic properties but a lower amount of critical rare-earth elements.
\end{abstract}

\maketitle

\section{Introduction}

Machine Learning (ML) is currently establishing itself as an important tool in materials research.
ML models have been successfully used for predicting grain boundary energies \cite{Kiyohara2016,Kiyohara2016b} and mobilities \cite{Rosenbrock2017} in pure metals, ferroelectric Curie temperatures and classes of perovskites \cite{Balachandran2016,Pilania2015,Yang2014}, and electronic \cite{Lopez-Bezanilla2014,Montavon2013} as well as magnetic properties \cite{Cheng2014,Dam2014,Kusne2014}.
The main applications of ML have, however, been limited to learning properties that are directly related to total/atomic energies or atomic forces \cite{Ghiringhelli2015,Faber2015,Faber2016,Balabin2011,Bartok2013,
Bartok2010,Li2015,Botu2015,Botu2015a,Oganov2009,Suzuki2017,Snyder2012,Pilania2013,Huan2015,Rupp2012}. We believe that this limitation originates mainly from the availability of high-accuracy density-functional theory (DFT) data for the energies of millions of structures and compositions in internet databases, such as the Materials Project \cite{Jain2013}, 
the Open Quantum Materials Database (OQMD) \cite{Saal2013,Kirklin2015}, or 
the Novel Materials Discovery (NOMAD) repository \cite{nomad}.
The potential of ML to accurately predict and to optimize other materials properties, 
which are less well documented in internet databases, needs thus to be further explored.

The search for new hard-magnetic phases is an ideal application case for demonstrating the benefits of ML in materials research.
On the one hand, many hard-magnetic materials are intermetallic phases that contain a multitude of atoms per unit cell. 
The unit cell of the prominent Nd$_2$Fe$_{14}$B phase, for instance, contains 68 atoms.
This makes it infeasible even for combinatorial high-throughput screening (HTS) to cover the entire variety of possible phase compositions.
For a reasonable variety of screened elements, the number of possible compositions and atomic arrangements within the unit cell easily adds up to billions or even trillions.
ML overcomes this limitation by building a \textit{continuous} model for the learned properties based on the training data.
This enables a reliable interpolation between the training data and an accurate prediction of properties for unknown compositions. 
Today's best hard-magnetic materials, which are key components in many green-energy technologies, contain substantial amounts of supply-critical rare-earth (\textsc{Re}) elements, such as Dy, Nd, or Sm.
There is thus a considerable industrial demand for new materials with good hard-magnetic properties but less \textsc{Re} contents.

The objective of this paper is therefore to construct ML models for hard-magnetic properties and to subsequently use them for compositional optimization in order to identify those chemical compositions that exhibit good magnetic properties, but contain only few \textsc{Re} elements.
For training the ML models, we select the ThMn$_{12}$-type crystal structure \cite{DeMooij1988} from our materials database of hard-magnetic phases \cite{Koerner2016} as a promising substitute for the \textsc{Re}-rich state-of-the-art materials Nd$_2$Fe$_{14}$B and SmCo$_5$.
Instead of using all the data in our database for training, we use a subset with equally-distributed and equally-spaced supporting points where only one alloying element (besides Fe) is considered.
We believe that such a scenario where one holds a data set with not much more than 200 different configurations is representative for a typical systematic density-functional-theory (DFT) study of a certain non-standard property (which in our case is the magneto-crystalline anisotropy).

The paper is organized as follows:
In section~\ref{sec:method} we describe our computational approach, i.e., how we train the ML models to the hard-magnetic properties of the intermetallic phases in our database.
In section~\ref{sec:results} we present the results for validating, testing, and optimizing the ML models.
These results are discussed in section~\ref{sec:discussion} with focus on the accuracy and reliability of the ML predictions and on the potential of the optimized compositions. 
The discussion also includes comments on the achievements and limitations of our approach, on possible pathways for future extensions, and on discrepancies between the DFT training data and experimental results. 
The paper is summarized in section~\ref{sec:summary}.

\section{Methodology}
\label{sec:method}

\subsection{Machine Learning}  
\label{sec:methods_ml}

The list of machine-learning (ML) models that have been used for predicting properties and behaviors of materials includes a variety of methods such as artificial neural networks \cite{Chugh2017,Behler2007,Chowdhury2016,Artrith2017}, genetic algorithms \cite{Giri2013,Chugh2017}, Gaussian processes \cite{Bartok2010,Artrith2017}, decision-tree and random-forest models \cite{Pilania2015,Chowdhury2016}, and kernel-based methods \cite{Rupp2012,Faber2015,Chowdhury2016,Faber2016}. 
Which of the different methods is best suited for a certain analysis usually depends on the type (e.g., scalar values or images) and amount (hundreds, thousands or millions of samples) of the available training data and the targeted application of the ML model (value prediction/regression or classification).

In this work, we make use of the kernel-based Support Vector Regression (SVR) method to construct numerical ML models for $K_1$, $\mu_0M$ and $E_\text{f}$ (together denoted as target properties $y$ in the following).
SVR is a nonlinear regression analysis that makes use of the so-called kernel trick.
In kernelized ML methods, the kernel function maps the input space, in which the target property $y$ is usually not a linear function of the feature vector $\mathbf{x}$, into a higher dimensional space where such a linear relationship may exist. 
SVR is based on the concept of Support Vector Machines (SVMs), which were originally developed in the 1960s \cite{Vapnik1963} for classification purposes.
In SVR models, the predicted value $y^\text{m}$ (superscript 'm' for model) for a feature vector $\mathbf{x}$ is determined as
\begin{eqnarray}
y^\text{m}(\mathbf{x}) = \sum_\text{i} w_\text{i} k(\textbf{x}_\text{i},\textbf{x})+b,
\label{eq:kernel_model}
\end{eqnarray}
where $w_\text{i}$ are the individual weights for each training vector $\mathbf{x}_\text{i}$, $k(\mathbf{x}_\text{i},\mathbf{x})$ is the kernel function, 
and $b$ is the constant intercept ('$b$' stands for bias).

The kernel $k(\mathbf{x}_\text{i},\mathbf{x})$ is typically of linear, polynomial or Gaussian (radial basis function, 'rbf') type \cite{sklearn}:
\begin{eqnarray}
k^\text{linear}(\mathbf{x}_\text{i},\mathbf{x}) 		& = & \mathbf{x}_\text{i}\cdot \mathbf{x} \text{,} \\
k^\text{poly}(\mathbf{x}_\text{i},\mathbf{x}) 		& = & \left[\gamma (\mathbf{x}_\text{i}\cdot \mathbf{x})\right]^\text{d}, \\
k^\text{rbf}(\mathbf{x}_\text{i},\mathbf{x}) 			& = & \exp \left[-\gamma \|\mathbf{x}_\text{i}-\mathbf{x}\|_2^2 \right].
\end{eqnarray}
Here, the subscript '2' indicates the Euclidean ($L_2$) distance and '$\cdot$' the inner product, d is the degree of the polynomial kernel, and
$\gamma$ represents the width of the respective kernel function.

SVR uses an $\varepsilon$-insensitive loss function \cite{Vapnik1995, Smola2004}, 
which means that optimizing a SVR model involves minimizing the weights $w_\text{i}$. 
This procedure leads to a flat evolution of $y(\mathbf{x})$ and inherently reduces the risk of over-fitting \cite{Witten2011} by simultaneously allowing for some larger deviations (up to a value of $\varepsilon$) between $y$ and $y^\text{m}$ for individual samples ('outliers').
In this paper, we use $\varepsilon=$ 0.1 T, 1.0 MJ/m$^3$, and 0.01 eV/atom for the ML models for $\mu_0M$, $K_1$, and $E_\text{f}$, respectively.
This choice can be considered as the targeted accuracy of the constructed models. 

Besides the choice of the kernel function (and its width $\gamma$), 
SVR models have an additional regularization parameter $C$, 
which determines the trade-off between the flatness of $y^\text{m}(\mathbf{x})$ and 
the amount up to which deviations larger than $\varepsilon$ are tolerated \cite{Smola2004}
(a higher $C$ allows higher possible values for the weights $w_\text{i}$).
As mentioned before, $\gamma$ determines the width of the kernel and 
therefore the contribution of neighboring feature vectors to 
the prediction (a smaller $\gamma$ resulting in a higher contribution of samples close by).
For a given kernel function, both hyperparameters control the complexity of the model and
need to be optimized in order to prevent under- and over-fitting.
To yield a model that is as universal as possible, both $C$ and $\gamma$ should be as small as possible.
Note that $\gamma$ has no meaning for a linear kernel.
                                   
Extensive descriptions of the SVR method are given in Refs.~\cite{Vapnik1995, Smola2004}.
The SVR algorithm is used in its implementations in the Python ML package \texttt{scikit-learn} \cite{Pedregosa2011,sklearn}.
To assess the potential benefit of ML over classical fitting methods, 
we also parameterized linear regression (LR) functions ($y^\text{m} =  \sum_\text{j} w_\text{j} x_\text{j}+b , j = 0\ldots N_\text{features}$) for our data. Note that in this case, contrary to the SVR method described above, the weights  $w_\text{j}$ are determined for each component of $\mathbf{x}$, i.e., for the feature $x_\text{j}$, not for each training sample $\mathbf{x}_\text{i}$.

\subsection{Material Database}
\label{sec:methods_database}

The material database used for training and validating the ML models originates from our previous study on rare-earth-lean intermetallic \textsc{ReA}$_{12}$X compounds \cite{Koerner2016}.
Here, \textsc{Re} is either Ce or Nd, and A can be one of the magnetic transition-metal elements Mn, Fe, Co, and Ni, the non-magnetic elements Ti, V, Cr, Cu, Zn, Al, Si, and P, or a mixture of them.
The element X is one of the small interstitial elements B, C, or N.

\begin{figure}[htb!]
\includegraphics[width=\columnwidth]{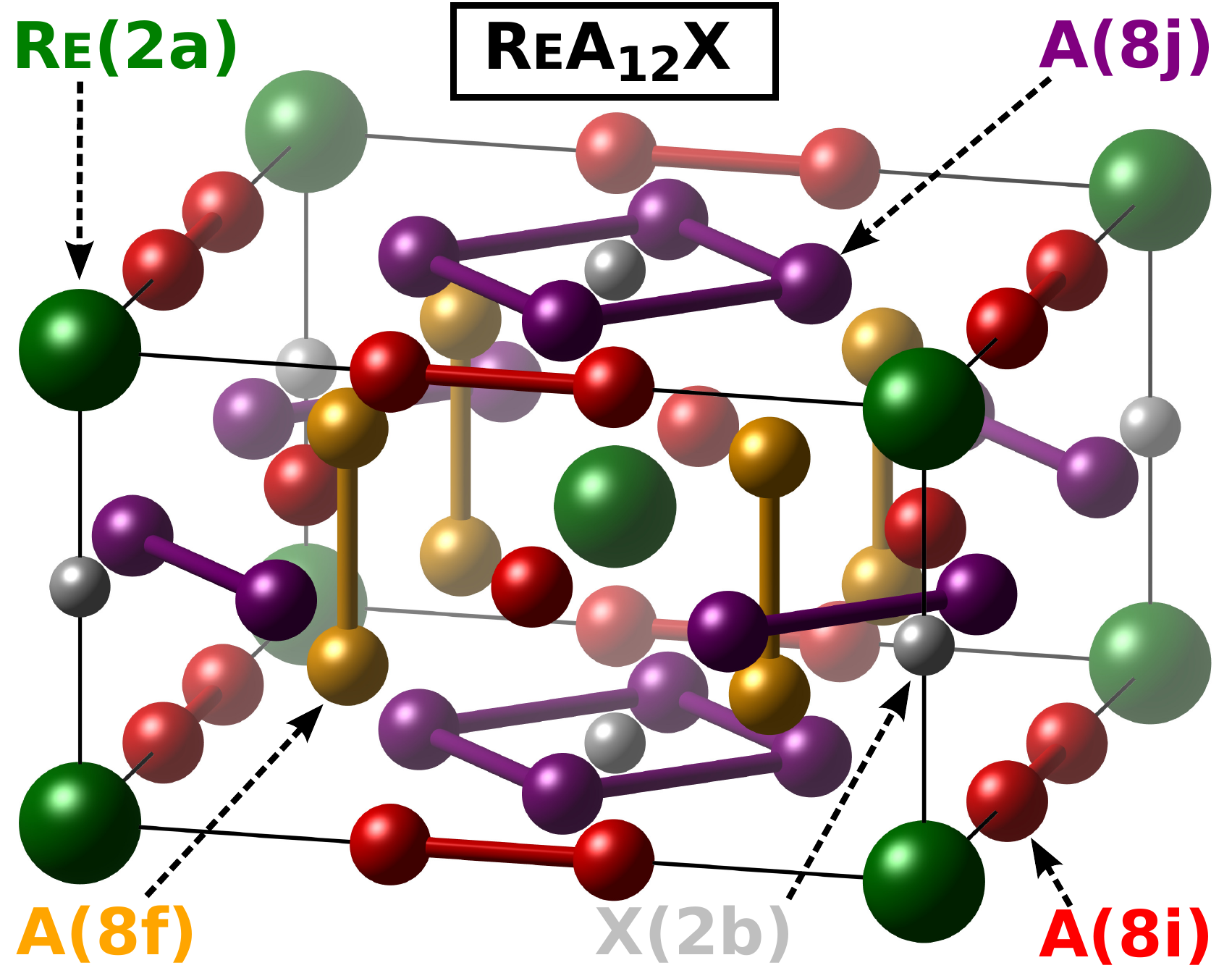} 
\caption{Crystal structure of the \textsc{ReA}$_{12}$X structure with indicated Wyckoff positions.
The bonds between symmetry-equivalent A atoms are drawn for better visualization.
\label{fig:structure}}
\end{figure}

The \textsc{ReA}$_{12}$X structure is based on the ThMn$_{12}$ type, 
which was first reported by de Mooij \etal{} \cite{DeMooij1988}.
The symmetry-equivalent Wyckoff sites of this body-centered tetragonal structure (space group \#139, I4/mmm symmetry) are the 2a (occupied by \textsc{Re} elements), 8i, 8j, 8f (occupied by A elements), and 2b (occupied by X elements, if any), see
Fig.~\ref{fig:structure} for illustration. 
Note that the number of atoms per formula unit (14) is only half of all atoms per conventional tetragonal unit cell (28).
The tetragonal lattice parameters $a=8.566$~\r{A} and $c=4.802$~\r{A}, as well as the crystal coordinates of the Wyckoff sites were taken from the work of Isnard \etal{} \cite{Isnard1998}.

To create the magnetic-properties database, we employ a HTS approach with DFT calculations using the tight-binding linear-muffin-tin-orbital atomic-sphere-approximation method (TB-LMTO-ASA) \cite{Andersen1975} based on the work of F\"ahnle \etal{} \cite{Hummler1992,Faehnle1993}.
The several thousand phases were calculated with the HTS setup described by Drebov \etal{} \cite{Drebov2013} which allows a fully automated generation of new phases by combinatorial substitution of sets of atoms sitting on specific positions.
In this study, for each combination of \textsc{Re} and X, all A positions (see Fig.~\ref{fig:structure}) are initially occupied with Fe atoms.
All Fe atoms sitting on a specific Wyckoff site are then substituted by other alloying elements A and this substitution is carried out for all combinations of Wyckoff sites.
To indicate this substitution schema, we refer to the generated compounds as \textsc{Re}(Fe,A)$_{12}$X in the following. Moreover, we have generated a substantial number of configurations where two alloying elements A and A' are considered and also allow for partial substitution on Wyckoff sites.

By means of DFT calculations, we determine (among other properties) the magnetization $\mu_0M$, the local magnetic moments, the uniaxial magneto-crystalline anisotropy constant $K_1$, and the relative phase-stability energy $E_\text{f}$.
Here, the latter is evaluated by comparing the total energy of a compound to the total energies of the elemental constituents (and not with all competing binary or ternary phases). Thereby, its computation is compatible with the approach of a high-throughput analysis and
it serves as a valuable first-order estimate for the expectable thermodynamic stability. 

The technologically interesting figures of merit of hard-magnetic materials, 
i.e., the maximal energy product $(BH)_\text{max}$ and the the anisotropy field $H_\text{a}$ are estimated from $\mu_0M$ and $K_1$ as follows \cite{Hirayama2015a}:
\begin{eqnarray}
(BH)_\text{max}^\text{EST} 	& = & \frac{(0.9\mu_0M)^2}{4\mu_0}, \label{eq:BH_max} \\
H_\text{a} 					& = & \frac{2K_1}{\mu_0M}. \label{eq:Ha}
\end{eqnarray}
Note that Eq.~(\ref{eq:BH_max}) implies the common assumption that ideally about 10\% of a processed bulk hard magnet consists of non-magnetic phases \cite{Hirayama2015a}.
For further details on the HTS approach using TB-LMTO-ASA calculations, we refer to Refs.~\cite{Drebov2013,Koerner2016}.

In order to convert our zero-temperature single-crystal, single-domain DFT-HTS results for $K_1$ and $H_\text{a}$ to room-temperature estimates, which can serve as guidelines for experimental efforts, the calculated values need to be divided by four for Nd. 
For Ce, a division by 35 provides a conservative estimate.
These heuristic factors have been derived from benchmark calculations
of well-known hard-magnetic materials (see Table II and discussion in Ref.~\cite{Koerner2016}).

For the optimization of the hyperparameters of our ML model (see Secs. \ref{sec:methods_ml} and \ref{subsec:buildingML}) we used the \textsc{Re}Fe$_\text{12-4z}$A$_\text{4z}$X (z~=~$0\ldots3$) subset of all the different \textsc{Re}(Fe,A)$_{12}$X configurations in our database.
In this data set, each A position (8i, 8j, 8f) which initially has been occupied by Fe atoms, was subsequently fully replaced by other alloying elements A.
For each choice of the $N_\text{X} = 3$ choices for X, there exists one configuration without alloying elements (i.e. z~=~0) and $N_\text{sub} = 7$ possible configurations for each of the $N_\text{A} = 11$ alloying elements A, namely three for both z~=~1 and z~=~2 and one for z~=~3. For \textsc{Re}~=~Nd, the data set therefore consists of 234 [$=N_\text{X} \cdot (1 + N_\text{sub} \times N_\text{A} ) = 3 \cdot (1+7\times11)$] configurations.
For \textsc{Re}~=~Ce, two compounds could not be converged in the given crystal structure, thus the data set contains only 232 configurations.

In total, our database contains 3,080 entries. Besides the 466 configurations used for training the ML models, the data includes 2,614 compounds with multiple substitutional alloying elements and partially chemically heterogeneous occupations of Wyckoff sites. 
For example the 8i position can be occupied by two Co atoms and six Fe atoms.
This remaining data set was used for the subsequent testing (see Sec. \ref{sec:results_testing}) of our model with the previously optimized hyperparameters.
Note that the test data are inherently difficult to be predicted with our realistic choice of training data as they contain lots of situations about which the trained model was not informed at all (i.e., two different alloying elements in the same configuration). 
The objective of our work is not to create the best possible model (for which we should obviously use the whole data set for training) but to reflect a realistic scenario and therefore use the typical outcome of a systematic DFT study in terms of number and distribution of the data points.
         
\subsection{Material Representation}

For a straightforward decoding of the structural and chemical information stored in the feature vectors $\mathbf{x}$, the construction of $\mathbf{x}$ needs to be bijective.
This means that a reversible mapping exists between $\mathbf{x}$ and the represented atomistic configuration.
We introduce here a descriptor that sufficiently fulfills this requirement. 
It contains the number of atoms $n^\text{(i)}_\text{j}$ of each chemical species \textit{i} resolved on each of the Wyckoff positions \textit{j}, 
\begin{equation}
\mathbf{x} =  \langle
n^\text{(1)}_\text{1},\ldots, n^\text{(1)}_\text{P}, 
n^\text{(2)}_\text{1},\ldots, n^\text{(2)}_\text{P}, 
\ldots, 
n^{(N)}_\text{1}, \ldots, n^{(N)}_\text{P} \rangle ,
\label{eq:descriptor_general}
\end{equation}
where \textit{N} is the number of different elements in the data set and 
\textit{P} is the number of different Wyckoff positions in the crystal structure.
When certain positions are occupied only by certain elements 
(e.g., the 2a site of the \textsc{ReA}$_{12}$X structure is occupied only by \textsc{Re} atoms)
the general form of \textbf{x} given in Eq.~\ref{eq:descriptor_general} can be simplified by leaving out those entries for which $n^\text{(i)}_\text{j}$ is zero throughout the whole data set.

\begin{figure}[t!]
\includegraphics[width=\columnwidth]{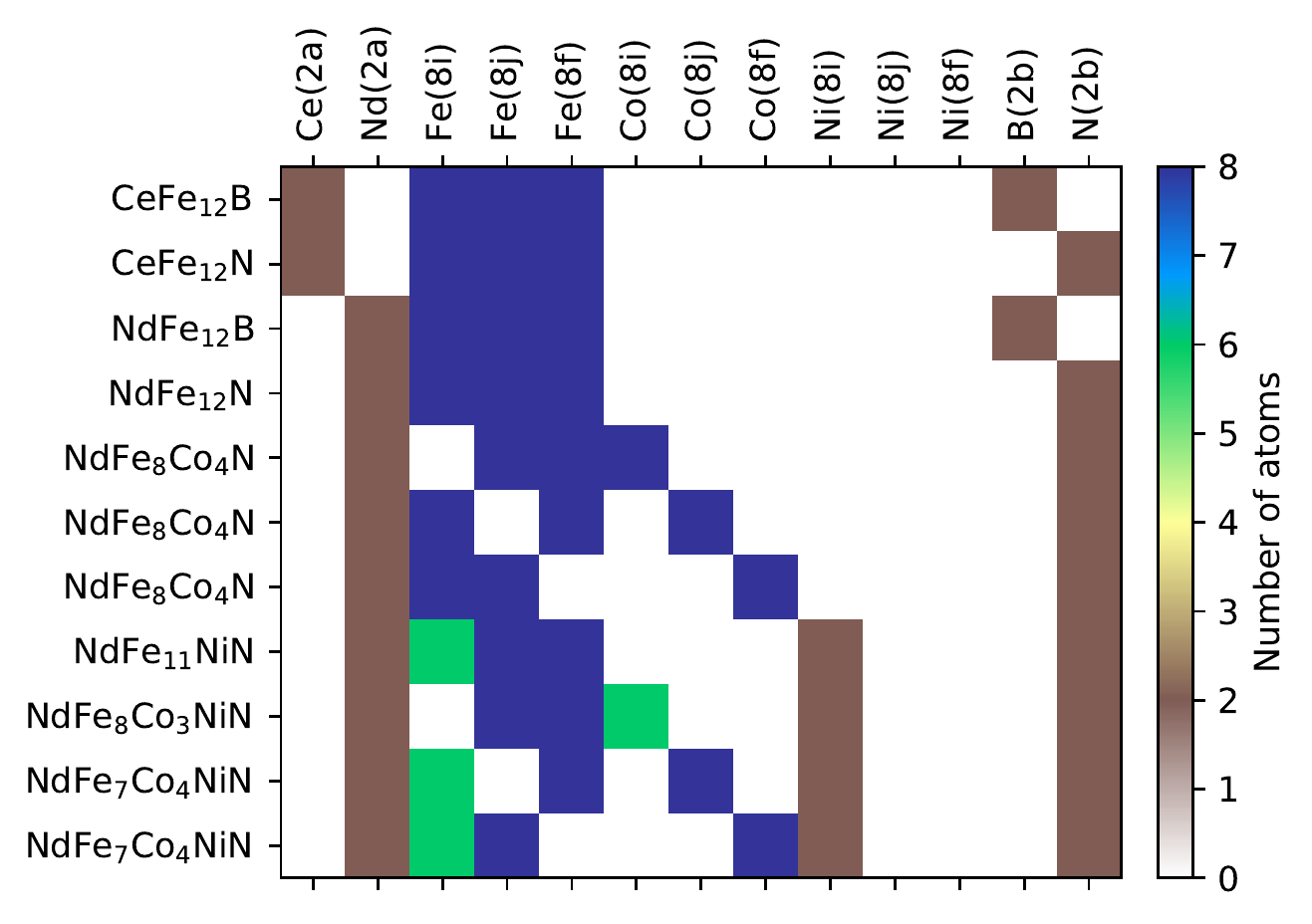} 
\caption{Illustration of descriptors for selected \textsc{Re}(Fe,A)$_{12}$X compounds. 
Note that the first seven configurations are included in the training set 
(\textsc{Re}Fe$_\text{12-4z}$A$_\text{4z}$X, z~=~$0\ldots4$)
whereas the last four compositions belong to the test set (see text for details).
\label{fig:Wyckoff_descriptor}}
\end{figure}

The descriptors for a data set of the \textsc{ReA}$_{12}$X structure with, for instance, two different \textsc{Re} elements on the 2a position ($j=1$), three different A elements on the 8i, 8j, and 8f positions ($j=2,3,4$), and two different interstitial elements on the 2b position ($j=5$) are then written as 
\small
\begin{equation}
\mathbf{x} = \langle 
n^\text{(1)}_\text{1}, n^\text{(2)}_\text{1}, 
n^\text{(3)}_\text{2}, n^\text{(3)}_\text{3}, n^\text{(3)}_\text{4}, 
n^\text{(4)}_\text{2},
\ldots ,
n^\text{(N-2)}_\text{4}, 
n^{(N-1)}_\text{5}, n^{(N)}_\text{5} \rangle 
\end{equation}
\normalsize
and have 13 components in total [instead of $N\cdot P = 7\cdot 5 = 35$ with Eq.~\eqref{eq:descriptor_general}].
Fig.~\ref{fig:Wyckoff_descriptor} illustrates the descriptors constructed in this way for selected \textsc{Re}(Fe,A)$_{12}$X compounds.

This description has the advantage that structural and chemical information are encoded separately.
We briefly note that this representation is somewhat similar to the descriptor which has been used very recently by Faber \etal{} for predicting the formation energies of elpasolite crystals \cite{Faber2016}. 

The construction of this descriptor involves a space group analysis of the crystal structure, 
for which we use the FINDSYM tool of Stokes \& Hatch as implemented in the ISOTROPY Software Suite \cite{isotropy,Stokes2005}.

\subsection{Optimization of Compositions}

To efficiently search for compositions that exhibit optimal combinations of $\mu_0M$ and $K_1$, we optimize the model functions $y^\text{m}(\mathbf{x})$ (with $y$ being $\mu_0M$ or $K_1$) using the stochastic basin-hopping optimization (BHO) algorithm 
as implemented in the Python package \texttt{scipy} \cite{scipy}. 
Details of the optimization procedure are given in the \ref{sec:optimization_details}.
BHO has the advantage of allowing both \textit{bounded} and \textit{constrained} components of $\mathbf{x}$. 
The components of the Wyckoff descriptor have to be \textit{bounded} because each of them has a lower limit (0) and an upper limit (maximal number of atoms per Wyckoff position, e.g. eight atoms for the 8i sites). 
At the same time, the components of \textbf{x} are \textit{constrained} because each non-zero component decreases the possible maximal value of all other components representing the same Wyckoff position.

After optimization, the descriptor $\mathbf{x}'$ that fulfills the search criteria, i.e., $y^\text{m}(\mathbf{x}') > y_\text{t}$ (subscript 't' for target), can be directly mapped to the corresponding chemical composition and distribution of the elements on the Wyckoff positions.
This fast and efficient procedure directly benefits from the bijective nature of the Wyckoff descriptor, which allows a reversible mapping between $\mathbf{x}$ and the respective material.

In the optimization, we effectively seek descriptors that maximize the anisotropy coefficient $K_1$ (and thereby the anisotropy field $H_\text{a}$) by using a relatively conservative target value for $\mu_0M_\text{t} = 1.4$~T since it is well known that the highest values for $\mu_0M$ are given by compositions containing mainly Fe with (small) contents of Co (see Slater-Pauling curve \cite{Slater1937,Bardos1969,Miyake2014}).
Note that we are not aiming at minimizing $E_\text{f}$ as a consequence of limitations in the input data since our reference database for determining $E_\text{f}$ contains only elemental crystal phases to compare with.
For this reason, we only use the criterion  $E_\text{f}^\text{m} < E_\text{ft}$ with a moderate target value $E_\text{ft}$ = 0.1 eV/atom for the optimized compositions (because $\pm$0.1 eV/atom is about the predictive power of the TB-LMTO-ASA method for compound formation energies).

\section{Results}
\label{sec:results}

Since the hyperparameters for the SVR algorithm (kernel function $k$, regularization parameter $C$, and width $\gamma$) are not known beforehand, we first determine the optimal set of hyperparameters that 
maximizes the predictive power of the model and minimizes its tendency for over-fitting.
The performance of the optimized ML models is then tested for unseen compositions from our database.
Finally, we use an optimization procedure to find the material composition or---more precisely---its descriptor, for which our two figures of merit are within the desired range of values: $\mu_0M > 1.4$~T and $K_1$ maximal.


\subsection{Training the ML Models}
\label{subsec:buildingML}


To find the optimal set of hyperparameters for the ML models, we varied the regularization parameter $C$ for the SVR as well as the kernel width $\gamma$ for various kernel functions. 
$C$ was varied between 0.1 and 1000 and $\gamma$ between 10$^{-7}$ and 1.0, with the kernels being linear, polynomial (degree d~=~2), and rbf functions. 
The numeric hyperparameters $C$ and $\gamma$ were varied on a logarithmic grid.

To validate the ML models built with a specific set of hyperparameters, 
we determine the Pearson correlation coefficient $\rho$ and the mean absolute error (MAE) \cite{Hastie2009,Witten2011},
\begin{eqnarray}
\rho(y,y^\text{m})		 	& = & \frac{\text{cov}(y,y^\text{m})}{\sigma_y \sigma_{y^\text{m}}} 
,
\label{eq:correlation_coefficient}\\
\text{MAE}(y,y^\text{m})	& = & \frac{\sum_{i=1}^N\| y^\text{m}-y\| }{N_\text{samples}},
\label{eq:mae}
\end{eqnarray}
where cov is the covariance function, $\sigma$ denotes the standard deviation, and $N_\text{samples}$ is the number of samples.
The correlation coefficient $\rho$ measures the linear relationship between two (normally distributed) datasets. 
It varies between $-1$ and $+1$, where values close to zero indicate that no correlation exists. 
The limiting values $\rho = +1$ ($-1$) imply a perfectly direct (indirect) linear proportionality between $y^\text{m}$ and $y$.
The MAE (in units of the predicted property) indicates the average difference between $y^\text{m}$ and $y$ and should therefore be as small as possible.

To estimate whether the model is over-fitted, i.e., it exactly matches the training samples but nothing else, we performed tenfold cross validation (CV) \cite{Witten2011}:
The data set used for building and optimizing the ML model consists of 234 (232) samples of type 
\textsc{Re}Fe$_\text{12-4z}$A$_\text{4z}$X for \textsc{Re}~=~Nd~(Ce). It is randomly divided into ten subsets of which nine are used to train a new ML model and the tenth is used for validation.
This procedure is repeated ten times and the average values of the correlation metrics $\rho$ and MAE for the ten validation sets are used as a measure of the performance of the model with respect to unseen test data. 

\begin{table}[htbp]                                                                                                             
\caption{\label{tab:results_validation_testing}
Optimal set of hyperparameters [kernel $k$, degree d for polynomial ($p$) kernels, regularization parameter $C$, and kernel width $\gamma$] and correlation metrics ($\rho$ and MAE) from tenfold CV.
$\mu_0M$ varies between 0 and 2 T;
$E_\text{f}$ varies between $-1$ and 1 eV/atom;
$K_1$ varies between 10 and 90 MJ/m$^3$ for Nd and between 50 and 230 MJ/m$^3$ for Ce.
For $\mu_0M$, only configurations with a significant magnetization of $\mu_0M > 0.25$ T have been used for training (181 for Nd, 179 for Ce).
}   
\footnotesize
\begin{tabular}{cccc}                                                                                                
\hline
\hline
Property & Hyperparameters & CV (Nd) & CV (Ce)\\
(unit) &  $k$(d), $C$, $\gamma$ & $\rho$ (MAE) & $\rho$ (MAE)\\
 \hline                                                                                                                   
$\mu_0M$ (T) 			& $p$(1), 100, --- & 0.95 (0.09)	& 0.91 (0.11) \\
$K_1$ (MJ/m$^3$) 		& $p$(2), 100, 0.1 	& 0.90 (3.9)	& 0.88 (10.3) \\
$E_\text{f}$ (eV/atom)	& $p$(2), 0.1, 0.1 	& 0.95 (0.043) 	& 0.94 (0.04) \\
\hline
\hline
\end{tabular}                                                                                                             
\end{table}

\begin{figure*}[t!]
\includegraphics[width=\textwidth]{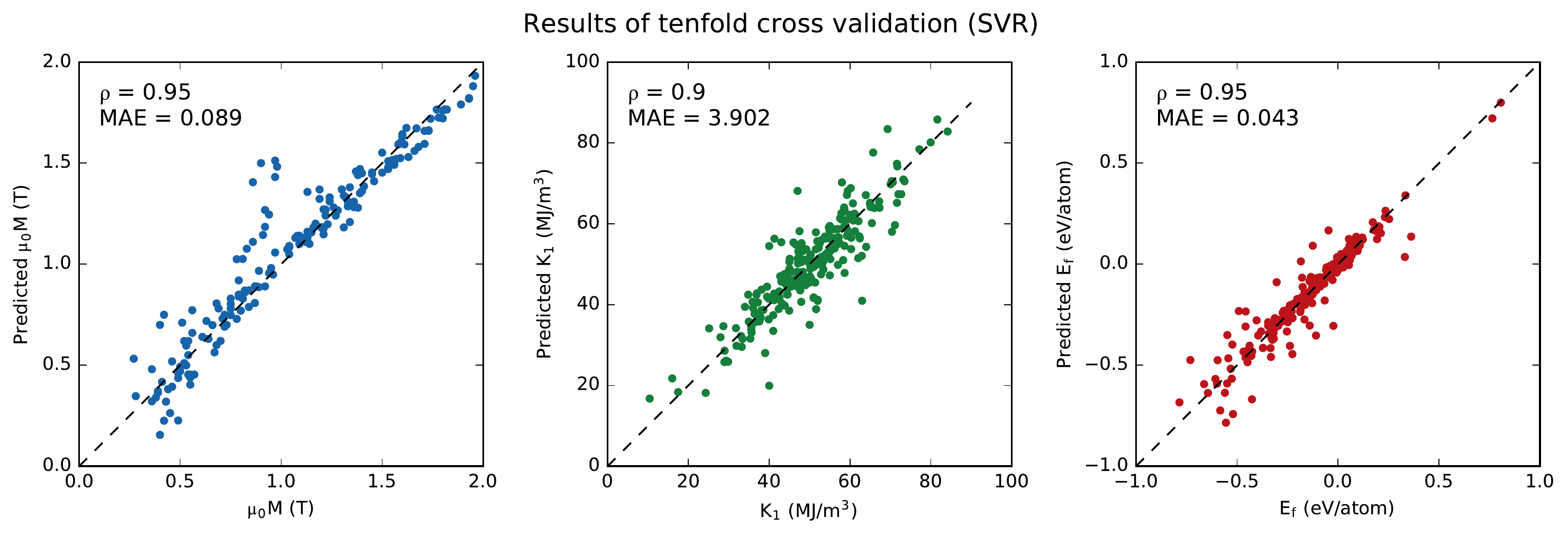} 
\caption{Results of the tenfold cross validation of the SVR models for $\mu_0M$ (left), $K_1$ (center), and $E_\text{f}$ (right) for Nd-containing compounds.
The colored circles indicate the predictions for each of the ten validation runs.
The used (optimal) sets of hyperparameters  are given in Tab.~\ref{tab:results_validation_testing}.
The MAEs are given in the units of the corresponding axes. 
\label{fig:results_validation}}
\end{figure*}

The optimal set of hyperparameters for each combination of ML model and material property were determined by taking those hyperparameters for which $C$ and $\gamma$ were minimal and satisfying $\rho > 0.95 \rho_\text{max}$ ($\rho_\text{max}$: maximal correlation coefficient for each model/property combination).
This ensured that the models were both as accurate and as universal as possible at the same time.
The optimal set of hyperparameters and the corresponding correlation metrics $\rho$ and MAE obtained with these hyperparameters are listed in Tab.~\ref{tab:results_validation_testing}.
The correlation coefficients obtained with CV are in the range of $\rho$~=~0.88$\ldots$0.95 for the models with the optimal set of hyperparameters.
The CV results are displayed in Fig.~\ref{fig:results_validation} for the Nd(Fe,A)$_{12}$X compounds.

\subsection{Testing of the ML models with unseen compositions}
\label{sec:results_testing}

\begin{figure*}[t!]
\includegraphics[width=\textwidth]{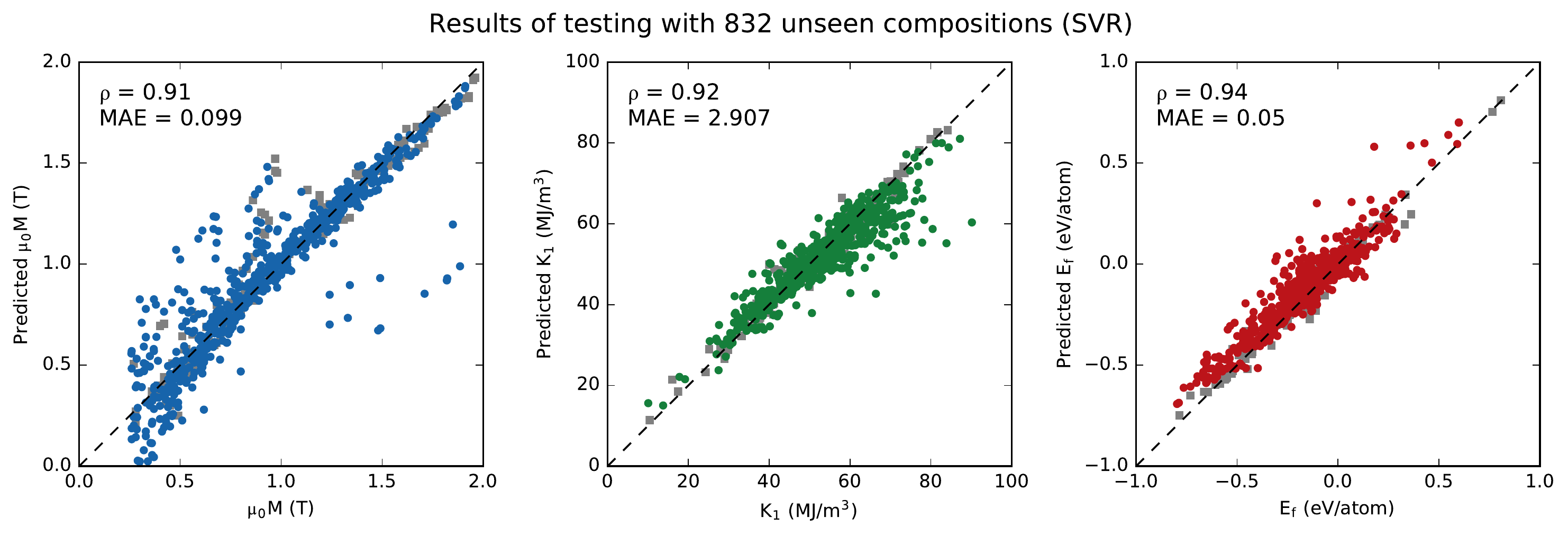} 
\includegraphics[width=\textwidth]{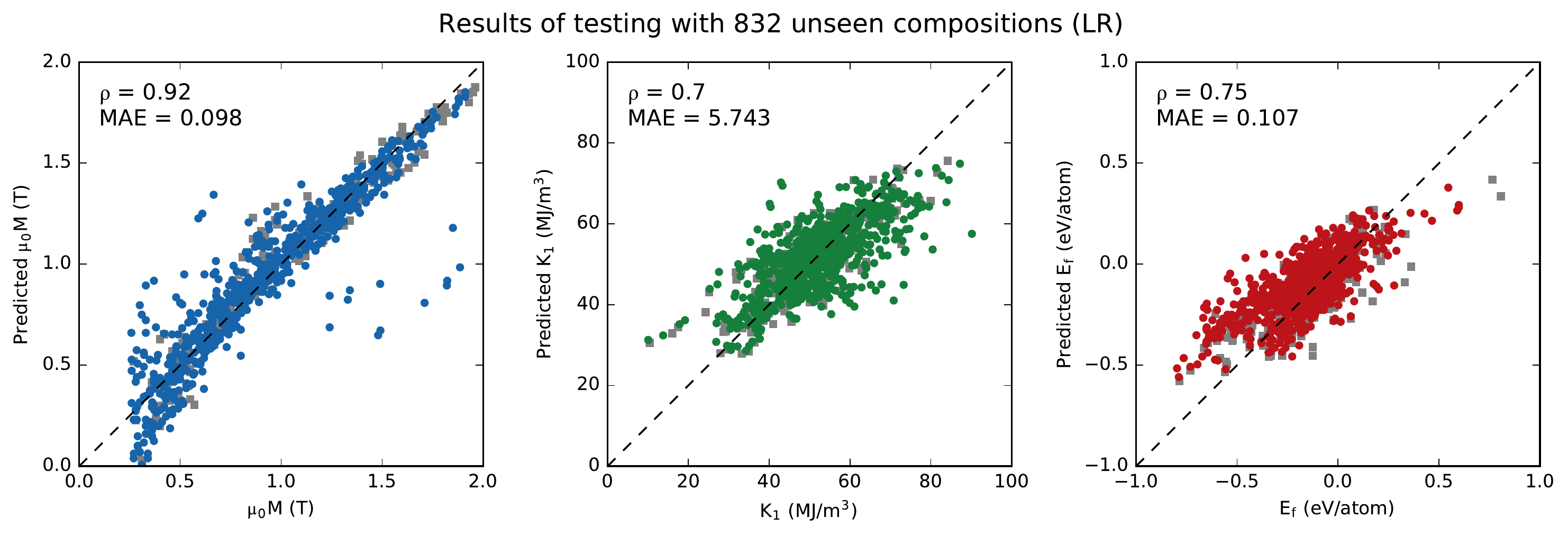} 
\caption{Testing of the SVR models (top) and LR (bottom) for $\mu_0M$ (left), $K_1$ (center), and $E_\text{f}$ (right) for Nd-containing compounds.
The colored circles indicate the corresponding predictions for the 832 unseen compositions used as test samples.
The grey squares are the predictions for the models trained on the whole data set (training and validation data).
The used (optimal) sets of hyperparameters for the SVR model are given in Tab.~\ref{tab:results_validation_testing}.
The MAEs are given in the units of the corresponding axes. 
\label{fig:results_testing}}
\end{figure*}

For testing our ML models we use test sets that contain configurations, where the the alloying atoms fractionally occupy the 8i Wyckoff site.
The resulting predictions for Nd(Fe,A)$_{12}$X are presented in Fig.~\ref{fig:results_testing}(top) for SVR and Fig.~\ref{fig:results_testing}(bottom) for LR.
The determined correlation coefficients $\rho$ and MAEs are summarized in the last column of Tab.~\ref{tab:results_validation_testing}.
For $K_1$ and $E_\text{f}$, $\rho$ is somewhat lower than the CV results 
while the correlation for $\mu_0M$ is similar to the one determined by CV.

Most of the outliers in the SVR predictions for $\mu_0M$, see Fig.~\ref{fig:results_testing}(top), originate from Mn-containing compounds.
This can be attributed to the magnetic behavior of Mn, which can change from ferromagnetic to anti-ferromagnetic magnetism depending on the precise atomic arrangement \cite{Butcher2017}.
For $K_1$ and $E_\text{f}$, most of the outliers are attributed to phases where the non-ferromagnetic elements Al, P, Si, Ti, and Zn represent more than 66 \% of all A elements in the compound. 
In both cases, the evolution of the calculated properties with the elemental concentrations is less systematic and predictions are inherently difficult.
It is important to note that the outliers are, however, uncritical for the final optimization because they do not lead to compositions that meet our search criteria.

The test results give us a good indication on how well the ML models will interpolate between the training data, or in other words, how well the models will predict the whole property space of the \textsc{ReA}$_{12}$X crystal structure (for the trained \textsc{Re}, A, and X elements).
Note that the achieved MAEs are of the same magnitude as the desired accuracy $\varepsilon$ of the SVR models (see section \ref{sec:methods_ml}).

\subsection{Optimal Compositions}

A selection of the obtained optimal compositions is presented in Tab.~\ref{tab:optimal_compositions} along with their respective values for $\mu_0M$ and $K_1$ and the expected values for $(BH)_\text{max}$ and $H_\text{a}$ estimated according to Eqs.~(\ref{eq:BH_max}-\ref{eq:Ha}).

Even though the test results (section \ref{sec:results_testing}) already indicate that the ML predictions should be accurate,
we verified the ML-predicted values with \textit{a posteriori} performed TB-LMTO-ASA calculations.
In addition to the optimal compositions, the magnetic properties of the most promising candidates drawn from our database \cite{Koerner2016} as well as the state-of-the-art benchmark materials SmCo$_5$ and Nd$_2$Fe$_{14}$B are presented for completeness.
The predicted and calculated values are compared to experimental data whenever available.
Note that the values for the formation energy $E_\text{f}$ are not given in the table explicitly, 
but they are negative for all configurations shown.

\begin{table*}[t]
\caption{
Selected compositions with high $K_1$ and $\mu_0M$ above 1.4 T found by optimizing the ML models built for the \textsc{ReA}$_{12}$X crystal structure.
Predictions of the ML models are indicated with a superscript 'm' and compared to \textit{a posteriori} performed DFT (TB-LMTO-ASA) calculations.
The Wyckoff positions of Fe substitutes are specified as superscripts (e.g., Co$^\text{1/8i}_{0.5}$ means that Co occupies one out of eight 8i positions, which is equivalent to 0.5 atoms in the formula unit).
The best candidates from the HTS \cite{Koerner2016} are marked with an asterisk ($^*$).
The calculated $K_1$ values are scaled (after the slash) for better comparability with experiments (see section \ref{sec:methods_database} for the heuristic scaling factors).
The macroscopic values for $(BH)_\text{max}^\text{EST}$ and $H_\text{a}$ are estimated according to  Eqs.~(\ref{eq:BH_max}-\ref{eq:Ha}).
The available experimental values for $\mu_0M$, $K_1$, $(BH)_\text{max}$, and $H_\text{a}$ are given in parentheses.
$E_\text{f}$ is negative for all listed compositions.
 \label{tab:optimal_compositions}}
\footnotesize
\begin{tabular}{l | c c | c c | c c}
\hline
\hline
\multicolumn{1}{c|}{Composition} & \multicolumn{2}{c|}{Predictions of ML Models} & \multicolumn{2}{c| }{TB-LMTO-ASA Calculations} & \multicolumn{2}{c}{Estimations of macroscopic values} \\
& $\mu_0M^\text{m}$ (T) & $K_1^\text{m}$ (MJ/m$^3$) &
$\mu_0M$ (T) & $K_1$ (MJ/m$^3$) & $(BH)_\text{max}^\text{EST}$ (kJ/m$^3$) &  $H_\text{a}$ (T)\\
\hline
CeFe$_6$Co$_4^\text{8/8f}$Cu$_{1.5}^\text{3/8i}$Ti$_{0.5}^\text{1/8i}$N &
1.45 & 165 & 1.45 & 169 / 4.8 & 337 & 233 \\
CeFe$_7$Ni$_4^\text{8/8f}$Co$^\text{2/8i}$N &
1.48 & 167 & 1.57 & 170 / 4.9 & 398 & 216 \\
CeFe$_{8}$Ni$_{4}^\text{8/8f}$N$^*$	& --- & --- & 1.61 	& 167 /	4.8 & 417 & 207 \\
NdFe$_{6.5}$Co$_{3.5}^\text{2/8i,5/8f}$P$_{1.5}^\text{3/8j}$Ti$_{0.5}^\text{1/8i}$C & 
1.42 & 57 & 1.46 & 60 / 15 & 346 & 83 \\
NdFe$_{8.5}$Cu$_{3.5}^\text{3/8i,4/8j}$N & 
1.48 & 57 & 1.57 & 55 / 14 & 397 & 70 \\
NdFe$_{8}$Ni$_{4}^\text{8/8f}$N$^*$	& --- & --- & 1.68 	& 57 / 14	& 457 & 67\\
\hline
\multicolumn{7}{c}{\textit{Benchmark Materials}}\\
\hline
SmCo$_5$ 				& --- & --- & 1.07 (1.07 \cite{Coey2010}) & 69 (26 \cite{Coey2010}) & 184 (219 \cite{Buschow2005}) & 129 (40.4 \cite{Coey2010})\\
Nd$_2$Fe$_{14}$B 		& --- & --- & 1.87 (1.86 \cite{Herbst1991}) & 19 (4.9 \cite{Coey2010}) & 563 (516 \cite{Buschow2005}) & 20 (6.6 \cite{Coey2010})\\
\hline
\hline
\end{tabular}

\end{table*}

The optimal compositions given in Tab.~\ref{tab:optimal_compositions} exemplify the versatility of our approach, i.e. the combination of an optimization algorithm with a bijective descriptor.
Two compositions were, for instance, optimized under the constraint that they should contain Ti,
which is known for its beneficial effect on the phase stability \cite{Buschow1991b}.
The predicted NdFe$_{8.5}$Cu$_{3.5}$N phase, as another example, 
avoids the elements Co and Ni, which are both more expensive than Cu.

\section{Discussion}\label{sec:discussion}
\label{sec:discussion}

In the discussion, we focus on the comparison between the ML predictions and DFT-determined values of the optimal compositions. 
In addition, we critically assess the potential of the proposed compositions to substitute the state-of-the-art materials NdFe$_{14}$B and SmCo$_5$. 
Finally, we demonstrate the potential benefit of using kernel-based ML methods over classical LR fitting.

\subsection{Comparison between ML predictions and DFT}

Although the ML models were trained only to samples containing one alloying element (in addition to Fe), it is remarkable that the predictions for the optimal compositions (which contain often at least two alloying elements) are in overall very good agreement with the DFT calculations performed \textit{a posteriori} (the average ML-DFT difference is below 4 \%).
A plausible reason for this is the property of iron, other transition metals (\textsc{Tm}), and their alloys (whose electronic band structures are mainly formed by the \textsc{Tm} d-orbitals) that along the \textsc{Tm} series in the periodic systems the band structures and densities of states for a specific crystal structure are only scaled in width but hardly changed in shape with varying number of electrons occupying the bands (see, e.g. Refs. \cite{Moruzzi1978,Pettifor1995}).

\begin{figure}[b!]
\includegraphics[width=\columnwidth]{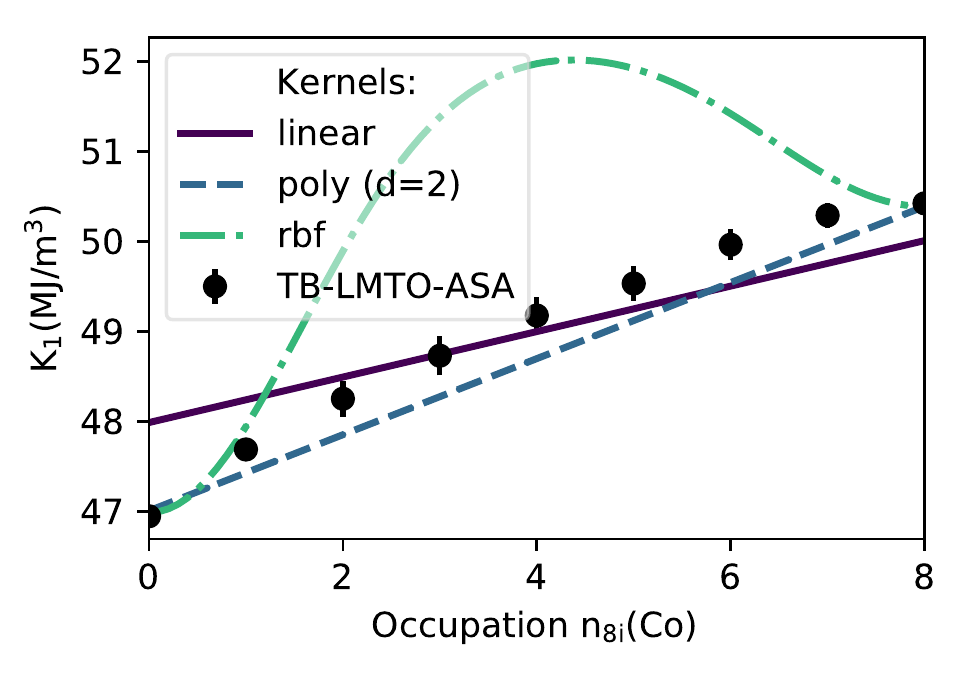} \caption{(Color online)
Influence of the kernel function on the prediction of $K_1$ for unseen compositions NdFe$_\text{12-z/2}$Co$_\text{z/2}$N (z~=~$n_\text{8i}\text{(Co)} = 0\ldots 8$) 
in comparison to TB-LMTO-ASA results.
The model was trained using the linear, polynomial (degree d~=~2), and 'rbf' kernels.
Only the values for $n_\text{8i}\text{(Co)}=0$ and 8 were used  for training.
\label{fig:results_kernelinfluence}}
\end{figure}

The apparent success of the machine to learn the underlying physics should not distract us from the fact that the interpolation behavior between training samples is mainly determined by the kernel function.
This property of ML models is visualized in Fig.~\ref{fig:results_kernelinfluence} where the predicted  $K_1$ values for different contents of Co on the 8i site, $n_\text{8i}\text{(Co)}=0\ldots 8$ (=~z), in the NdFe$_\text{12-z/2}$Co$_\text{z/2}$N  phase space are plotted for different kernel functions in comparison with TB-LMTO-ASA results.
Whereas with the polynomial kernel (d~=~2), the nearly linear increase of $K_1$ is very well reproduced, the linear kernel fails already in predicting the training data at z~=~0 and 8, respectively.
Due to its higher flexibility, the 'rbf' kernel exactly reproduces the training data points.
The drawback of this flexibility is that the deviation from the DFT results becomes very large in the range of z values.

One of the recipes of success for ML of materials properties is therefore the combination of a descriptor, whose components have an almost linear relationship with the modelled properties,
and a comparably inflexible kernel function.
Since many physical properties have polynomial relationships of low degrees with the material composition, 
this approach is expected to work equally well for other materials properties.

At the same time, one limitation of our approach is that the employed descriptor encodes all chemical and structural information implicitly.
ML Models built with this descriptor are thus not transferable to other crystal structures.
Promising \textit{universal} descriptors, which may overcome this shortcoming, have already been proposed for molecular systems \cite{Rupp2012,vonLilienfeld2015} and their suitability for use needs to be tested for periodic crystals in the future \cite{Faber2015}. 

\subsection{Potential of the optimized compositions}

The hard-magnetic properties of both benchmark materials, which contain comparably high amounts of \textsc{Re} elements, are currently known as the upper limits for $(BH)_\text{max}$ (NdFe$_{14}$B) and $H_\text{a}$ (SmCo$_5$). 
As to be expected, our candidates do not beat these values, but they represent alternatives with significantly reduced \textsc{Re} contents.


In comparison with the best results from the HTS (marked with an asterisk in the table),
the $K_1$ and $H_\text{a}$ values of most of the compositions found by optimizing the ML models
are similar or higher.
The largest improvement is achieved for NdFe$_{6.5}$Co$_{3.5}$P$_{1.5}$Ti$_{0.5}$C, for which $H_\text{a}$ is more than 20 \% higher than for the best candidate from the HTS in Ref.~\cite{Koerner2016}.
In the future, it would also be interesting to compare our best candidates obtained with BHO to predictions using multi-objective optimization, see Ref. \cite{Chugh2017} for an example in materials processing.

Finally, intermetallic phases with hypothetically good hard-magnetic properties can exist in reality only if they have sufficient thermodynamic stability. As stated above, in our DFT database of Ref.~\cite{Koerner2016} on which our present ML work is based, the formation energies of intermetallic phases are calculated with the TB-LMTO-ASA method and only with respect to their constituing single-component phases. The ASA is a well suited and justified approximation for calculating hard-magnetic key quantities $\mu_{0} M$ and $K_{1}$ of \textsc{ReA}$_{12}$-based phases with approximately close-packed crystal structure \cite{Koerner2016,Faehnle1993}. For estimating formation-energy differences between competing multi-component phases more precisely, total-energy calculations with more accurate {\em full-potential} DFT methods are necessary. However, theoretically predicted phases with promising hard-magnetic properties are already valuable information for experimentalists who want to produce hard-magnetic materials by metallurgical means even without prior theoretical stability prediction.  Future work on extending DFT databases to hard-magnetic phases with accurate phase-stability information is a desirable task. However, for the present ML work it is not an issue because the presented ML approach to predict chemical compositions of the specific \textsc{ReA}$_{12}$ structure type with good hard-magnetic properties, is not limited by a limited precision of the thermodynamic-stability data. Once that the DFT database is extended with more accurate {\em full-potential} DFT data for formation energies of competing multi-component phases, the presented ML approach will be immediately applicable to such an improved database.

\subsection{Comparison between ML and LR}

As it can be seen from the testing results in Fig.~\ref{fig:results_testing}, the $\mu_0M$ correlation metrics for the SVR ML model (with a linear kernel function) are nearly identical to those of LR.
This clearly illustrates that the benefits of ML come only into play if a non-linear kernel (as for $K_1$ and $E_\text{f}$) is used. In our work, this improvement is a factor of 2 for the MAE and around 0.2 for the correlation coefficient. 

\section{Summary}\label{sec:summary}
\label{sec:summary}

In this work, we successfully applied a ML approach to a common materials science problem, namely the search for a material composition that optimizes a certain physical property.
Based on our HTS database, we trained ML models that reliably predict the hard-magnetic key properties of \textsc{Re}(Fe,A)$_{12}$X compounds with up to three different non-ferrous alloying elements A based on input data with compounds containing only one alloying element.
By optimizing the ML model functions we identified optimal compositions for the \textsc{Re}A$_{12}$X structure that exhibit an increase in magneto-crystalline anisotropy field $H_\text{a}$ by more than 10 \% as compared to our previous HTS results.
Although in both $(BH)_\text{max}^\text{EST}$ and $H_\text{a}$ the predicted optimal compositions do not excel the current state-of-the-art materials SmCo$_5$ and Nd$_2$Fe$_{14}$B at the same time, 
they can still be valuable for bridging the gap between these [comparably low $(BH)_\text{max}^\text{EST}$ and high $H_\text{a}$ for SmCo$_5$ and vice versa for Nd$_2$Fe$_{14}$B] while having significantly lower \textsc{Re} contents and being therefore more cost efficient and less supply critical.


Our results clearly highlight the potential of ML methods for materials discovery and design.
With an appropriate choice of model hyperparameters and material representation it is possible to accurately predict the materials properties in the entire compound space and to identify compositions that optimize the learned property.
In the end, the applicability of ML is only limited by the availability and accuracy of training data.
For the future, we believe that ML of materials properties will be especially helpful, 
when the underlying structure-composition-property relationships are not yet completely clarified, as e.g., in the case of $K_1$ for novel \textsc{Re}-lean magnets.


\section{Acknowledgments}

Financial support for this work was provided by 
the Fraunhofer Lighthouse Project \textit{Critical Rare Earths}.

\appendix

\section{Optimization details}
\label{sec:optimization_details}

\begin{figure}[t!]
\includegraphics[width=\columnwidth]{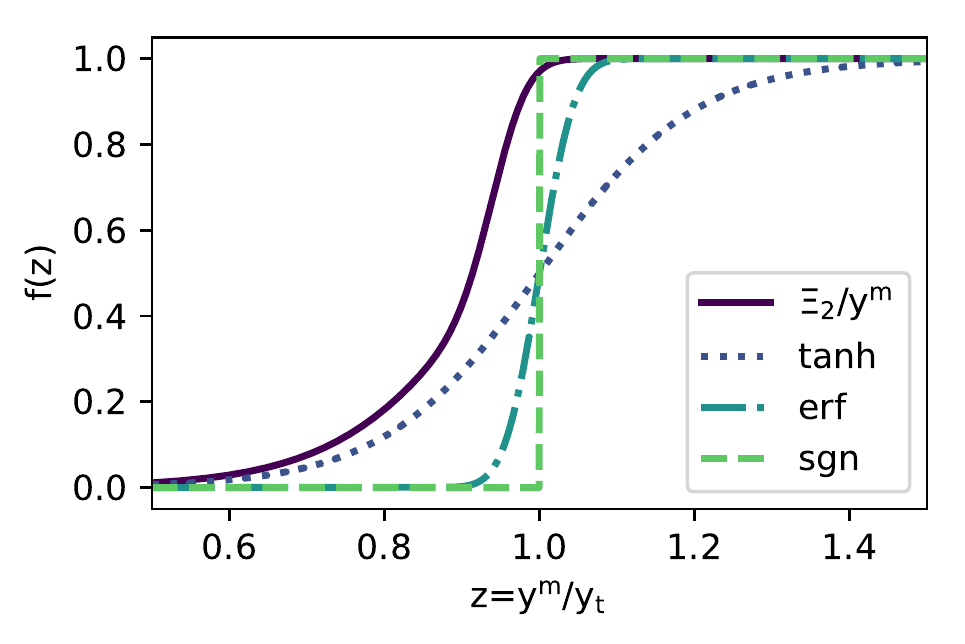}
\caption{(Color online)
Qualitative comparison of the asymmetric, smooth threshold function $\Xi_2(z)/y^\text{m}$ with
the point-symmetric $\tanh$, erf, and sgn functions where $z=y^\text{m}/y_\text{t}$.
Note that the arguments and parameters of the tanh, erf, and sgn functions were adjusted here to the threshold value 1 and 
the range of function values between 0 and 1 in order to be directly comparable with $\Xi_2$;
furthermore, for the tanh and erf functions the slope was adjusted in a similar manner as for $\Xi_2$, cf. Eq. \eqref{eq:xi2}.
\label{fig:results_thresholdfunctions}}
\end{figure}

Using basin-hopping optimization, we intend to determine the descriptor $\mathbf{x}$ that minimizes the following function:
\begin{equation}
F(\mathbf{x})=-\Xi_1(E_\text{f}^\text{m},E_\text{t}) 
\Xi_2(\mu_0M^\text{m},\mu_0M_\text{t}) \Xi_2(K_1^\text{m},K_\text{1t})
\label{eq:optfun}
\end{equation}
where the subscript 't' denotes the desired target values.
Note that $\mu_0M^\text{m}$, $K_1^\text{m}$, and $E_\text{f}^\text{m}$ are functions of $\mathbf{x}$, which is omitted in Eq.~(\ref{eq:optfun}) for better readability.
The smooth threshold function $\Xi_1$ ($\Xi_2$) becomes one ($\mu_0M^\text{m}$, $K_1^\text{m}$) below (above) the target value $E_\text{ft}$ ($\mu_0M_\text{t}$, $K_\text{1t}$). 
Both functions converge to zero otherwise.
The functions $\Xi_1$ (for $E_\text{f}$) and $\Xi_2$ (for $\mu_0M$ for $K_1$) are defined as follows:
\begin{eqnarray}
\Xi_1(z) &  = & \frac{1}{2}\text{erfc}\left[5(z-\frac{1}{2})\right]\label{eq:xi1}\\
\Xi_2(z)/y^\text{m} &   = &  1-
\frac{1}{4}\{\tanh\left[-5(z-0.95)\right]+1\} \nonumber \\ 
& & \{\text{erf}\left[-20(z-0.95)\right]+1\}\label{eq:xi2}
\end{eqnarray}
with $z=y^\text{m}/y_\text{t}$ and $y$ being either $\mu_0M$, $K_1$, or $E_\text{f}$, 
erf is the error function and erfc is its complement.
Obviously, the complementary error function in $\Xi_1(z)$ appears as being a logical choice for a function that should penalize values which become higher than certain target values.  

The functional form of $\Xi_2(z)$, on the other hand, needs some more explanation. 
It was chosen such that values smaller than the target value are less strongly penalized than values above it.
By comparing $\Xi_2$ with its two main point-symmetric constituents, namely the tanh and erf functions, this behavior is visually exemplified in Fig.~\ref{fig:results_thresholdfunctions}.
The somewhat unorthodox choice of $\Xi_2(z)$ has the objective to 'guide' the optimization algorithm in the correct direction (increasing $\mu_0M$ and $K_1$ values). 
Above the thresholds, however, all values contribute equally to $F(\mathbf{x})$, i.e., only by their magnitudes.

\bibliography{}

\end{document}